\documentclass[twocolumn]{aastex701}

\begin{document}

\title{Active Galactic Nuclei–driven Metallicity Enrichment in the Interstellar Medium of Mrk~573 }

\author[0000-0002-3626-5831]{D.~{\L}.~Kr\'{o}l}
\affiliation{Center for Astrophysics $|$  Harvard \& Smithsonian, 60 Garden Street, Cambridge, MA 02138, USA}
\affiliation{Astronomical Observatory of the Jagiellonian University, Orla 171, 30-244 Krak\'{o}w, Poland}
\email[show]{dominika.krol@cfa.harvard.edu}
\author[0000-0002-1333-147X]{P.~Zhu}
\affiliation{Center for Astrophysics $|$  Harvard \& Smithsonian, 60 Garden Street, Cambridge, MA 02138, USA}
\email{fake@example.com}
\author[0000-0002-3554-3318]{G.~Fabbiano}
\affiliation{Center for Astrophysics $|$  Harvard \& Smithsonian, 60 Garden Street, Cambridge, MA 02138, USA}
\email{fake@example.com}
\author[0000-0001-5060-1398]{M.~Elvis}
\affiliation{Center for Astrophysics $|$  Harvard \& Smithsonian, 60 Garden Street, Cambridge, MA 02138, USA}
\email{fake@example.com}
\author[0000-0001-8152-3943]{L.J.~Kewley}
\affiliation{Center for Astrophysics $|$  Harvard \& Smithsonian, 60 Garden Street, Cambridge, MA 02138, USA}
\email{fake@example.com}
\author[0000-0002-8659-3729]{N.~Murray}
\affiliation{Canadian Institute for Theoretical Astrophysics, University of Toronto, 60 Saint George Street, Toronto, ON M5S 3H8, Canada}
\affiliation{Canada Research Chair in Astrophysics, Canada}
\email{fake@example.com}
\author[0000-0001-9815-9092]{R.~Middei}
\affiliation{Center for Astrophysics $|$  Harvard \& Smithsonian, 60 Garden Street, Cambridge, MA 02138, USA}
\email{fake@example.com}
\author[0000-0001-8112-3464]{A.~Trindade~Falcão}
\affiliation{ NASA-Goddard Space Flight Center, Code 662, Greenbelt, MD 20771, USA}
\email{fake@example.com}

\begin{abstract}

We present the first spatially resolved { at $\sim20$~pc scale} application of AGN-specific metallicity diagnostics for the nearby Compton-thick Seyfert~2 galaxy Mrk~573 ($z = 0.017$) . We use Hubble Space Telescope narrow-band imaging, MUSE integral-field spectroscopy and apply { AGN strong-line metallicity diagnostics} based on [O~III], [S~II], H$\beta$, H$\alpha$, and [N~II] emission lines. We construct maps of $12 + \log \mathrm{(O/H)}$ for two different metallicity calibrations and two different N/O–O/H scaling relations out to $\sim1$~kpc and down to $\sim20$~pc scales. Our analysis reveals metallicity enhancement in AGN-dominated regions, with oxygen abundances reaching up to { few times Solar.}The metallicity shows a patchy spatial distribution, varying on $\sim100$~pc scales,  appears to trace { the high Seyfert/LINER index (SLI) value regions and }the VLA 6~cm jet/radio lobe emission.{ These spatial correspondences and  }  the lack of evidence for star formation in the bicone region suggest that the enrichment originates from metals transported from the nuclear AGN regions by winds, outflows, or jets. 

\end{abstract}


\section{Introduction} 

The metallicity of the interstellar medium (ISM) { is usually associated with the chemical evolution} of galaxies' stellar populations \citep[][]{Maiolino19,Kewley19}.
H~II region metallicity diagnostics applied to large spectroscopic surveys revealed the stellar mass-metallicity relation, and showed that low-mass galaxies are more metal-poor and prone to metal loss, while massive systems retain or recycle their metals \citep{Tremonti04}. Studies of individual galaxies have revealed negative radial {\rm O/H} gradients with a universal slope \citep{Ho15}.

The extent of Active Galactic Nuclei (AGN) influence on the chemical evolution of galaxies remains an open question. Several observations and simulations suggest that AGN activity can transport metals, either through jets \citep[][]{Kirkpatrick09} or outflows \citep{Du14, Amiri24}, out to $\sim10$~kpc \citep{Martin24}. Cosmological simulations indicate that jets and outflows can displace metal-rich gas from galaxies into the outskirts of clusters \citep{Fabjan10}. { \cite{Armah24} showed an anti-correlation between the metallicity and Eddington ratio  in the sample of nearby Seyfert galaxies. }However, other observational studies probing radial metallicity profiles find no clear evidence of AGN impact \citep{Amiri25}, and studies of MaNGA Seyfert sample revealed systematic nuclear under-enrichment relative to the disk metallicity\citep{doNascimento22}.
{ Overall, the metallicity of AGN has been studied less extensively than that of star-forming galaxies. Only a few optical strong-line metallicity diagnostics were available for AGN, including theoretical calibrations \citep{StorchiBergmann98} and semi-empirical calibrations \citep{Castro17,Carvalho20}. The mass-metallicity relation in AGN host galaxies has been investigated, but it exhibits larger scatter than that observed in star-forming systems \citep[][]{Thomas19,Oliveira24}, leaving open questions regarding the overall impact of AGN feedback on galactic chemical evolution.}

In this paper, we leverage the new { theoretical strong-line metallicity diagnostics } suitable for photoionization dominated (AGN-like) regions introduced by \citet[][Z24]{Zhu24}, and combine them with Hubble Space Telescope ({\it HST}) and MUSE data to study AGN-driven metal enrichment of the ISM. { As no three-dimensional (3D) AGN photoionization models currently exist, the Z24 metallicity diagnostics are calibrated using  one-dimensional (1D) AGN photoionization models of \cite{Zhu23}. While 1D models cannot capture the full anisotropic structure of the ISM, they have been shown to reproduce observed AGN emission-line spectra with reasonable accuracy and yield consistent predictions across UV, optical, and infrared wavelengths \citep{Zhu23}.} We present a detailed metallicity mapping of AGN-type regions in Mrk~573. Mrk~573 is a Compton-thick AGN \citep[absorbed by the equivalent column density $>1.6\times10^{24}$~cm$^{-2}$][]{Guainazzi05} hosted by a Seyfert~2 galaxy at $z=0.017$ \citep[scale $\sim 0.35$~kpc/'',][]{Springob05}. The extended emission in Mrk~573 has been studied across wavelengths, from deep X-ray {\it Chandra} observations of hot plasma emission \citep{Paggi12,Bianchi10}, through {\it HST} and {\it Gemini} optical and infrared observations of ionized gas \citep{Fischer10,Fischer17}, down to {\it VLA} radio observations of the jet and radio lobes at $6$~cm \citep{Ulvestad84}.

Using optical emission line diagnostics, we categorize Mrk~573 Narrow Line Regions (NLR) as Seyfert(AGN)- or LINER-type dominated. Then, based on two different metallicity diagnostics (Z24) we map the ISM metallicity up to $\sim1.0$~kpc from the nucleus, resolving $\sim20$~pc scales for AGN-type emission. We show that the metallicity is enriched in the ionization bicone regions, and $\log({\rm O/H})$ { values trace $6$~cm radio contours, and  
the ISM excitation using the recently developed Seyfert/LINER index \citep[SLI, ][]{Krol25}.} In Sec.~\ref{sec:data} we introduce the data sets and analysis  methods, and present the { VO \citep[Veilleux \&  Osterbrock,][]{Veilleux87} diagnostic diagrams}
and map for Mrk~573. In Sec.~\ref{sec:results} we  present the metallicity maps. In Sec.~\ref{sec:discussion} we discuss potential sources of metal enrichment. 
In Sec.~\ref{sec:summary} we summarize our conclusions.

Through the paper, we use $\Lambda$CDM cosmology with $H_0=69.3$~kms$^{-1}$Mpc$^{-1}$ and $\Omega_m=0.287$ \citep{Hinshaw13}.

\section{Data and analysis}\label{sec:data}

\subsection{Data}

We used archival {\it HST} observations in key diagnostic lines, [O~III]$\lambda5007$, H${\alpha}$+[N~II]$\lambda\lambda6548,83$, H${\beta}$, and [S~II]$\lambda\lambda6717,31$\footnote{Data set \dataset[DOI:10.17909/ksh5-7168]{https://doi.org/10.17909/ksh5-7168}}, in the narrow-band filters FR533N, FR680N, FR656N, and FR505N, respectively, characterized by $0.05^{\prime\prime}$ resolution (see \citealt{Ma21} for details).

To disentangle the fluxes of the H${\alpha}$ and [N~II] emission lines we used { the Very Large Telescope/MUSE integral field unit (IFU) data. Mrk~573 was observed with MUSE on January 24th 2021, with $72$~min total exposure time on the wavelengths covering the range of $4650$-$9300$\AA \citep{Bacon10}. Observations were carried out for the $1.53^{\prime}\times1.53^{\prime}$ field of view, with $0.2^{\prime\prime}$ spatial resolution  and seeings' full width half maximum of $1.593^{\prime\prime}$. For the data analysis we followed \citet{Zhu25} approach, and performed single Gaussian fits to the emission lines using the penalized pixel-fitting routine module \citep{Cappellari04,Cappellari17} incorporated in the nGIST pipeline \citep{Bittner19}\footnote{http://ascl.net/1907.025}. }

\subsection{VO mapping}

The distinction between star-forming, AGN-dominated, and low-ionization nuclear emission-line region (LINER) dominated galaxies is based on the BPT \citep[Baldwin, Phillips \& Terlevich,][]{Baldwin81,Kewley06} { and VO }diagrams \citep[][]{Veilleux87}. Using spatially resolved data,  { BPT/VO} maps can be built to determine the dominating driver of line emission in different regions within the same galaxy \citep[][]{Maksym16,Ma21}. { We used here { VO} diagrams/maps which are based on a comparison of the [O~III]/H$\beta$ and [S~II]/H$\alpha$  flux ratios}.  With the {\it HST} data, such analyses can be performed at $\sim0.1^{\prime\prime}$ scales \citep{Maksym16}, corresponding to the linear scales of  tens of parsecs in the nearest AGN \citep{Ma21,Maksym21,Trindade25}.

In the BPT { and VO} diagrams AGN-type (Seyfert-type) emission is typically attributed to AGN photo-ionization excitation, though {it} can also be produced by fast shocks with a photo-ionizing precursor \citep[][]{Allen08}. { The LINER-type emission can be associated with a variety of mechanisms, including post-asymptotic giant branch stars \citep{Binette94,Stasinska08}, low-luminosity/obscured AGN radiation \citep{Halpern83,Kraemer08} and shock excitation \citep{Halpern83,Dopita95}. The last two are the most likely in the central regions of Seyfert galaxy \citep{Ma21}}.

\subsection{SLI mapping}

The { VO} mapping can be extended to spatially probe the local small-scale variations of  the ISM excitation, rather than to rely on a simple LINER/Seyfert dichotomy. SLI is calculated on a pixel-by-pixel basis as the Seyfert and LINER type pixel perpendicular distance to the Seyfert/LINER division line \citep{Krol25}.

\begin{figure*}
    \centering
    \includegraphics[width=1\textwidth]{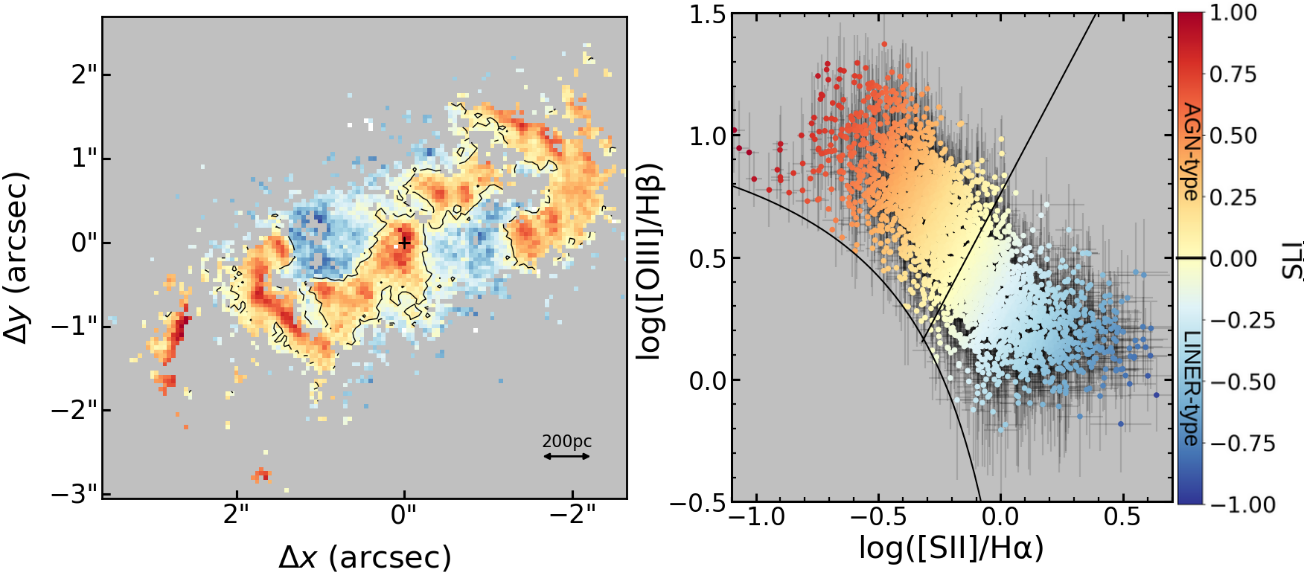}
     \caption{SLI map and diagram for Mrk~573. Left panel: SLI map, with colors reflecting SLI values, i.e. the distance of each pixel from the Seyfert/LINER division line in the { VO} diagram. Black contours trace points on the Seyfert/LINER division line (SLI$=0$). The black ``X'' marks the position of the nucleus. Right panel: corresponding VO diagram, with red indicating Seyfert-like excitation and blue denoting LINER-like excitation. The color intensity scales with the SLI value.}
    \label{fig:SLI}
\end{figure*}

We constructed the { VO} { diagram} for Mrk~573 { and separated AGN- and LINER-type emission} by comparing the values of $\log$([O~III]/H$\beta$) and $\log$([S~II]/H$\alpha$) \citep[][]{Kewley06}. Fluxes were calculated pixel-by-pixel, only for pixels with narrow-band filter fluxes at or above the $3\sigma$ level, where $\sigma$ is the standard deviation of the pixel count in a background-dominated region \citep[][]{Ma21}. { We color-coded both the map and VO diagram based on the SLI value. }The uncertainties are derived using the standard error propagation formula for uncorrelated variables \citep[see Appendix B in][for details]{Krol25}, assuming that $\sigma$ represents the uncertainty for each individual filter observation. The SLI map and the corresponding diagram in Fig.~\ref{fig:SLI} show more structure than a simple Seyfert/LINER map. We color-code SLI values on both the { VO} map and the diagram. {By construction,} LINER points (in blue) have SLI$<0$, and Seyfert points (in red) have SLI$>0$. The black contour in the left panel represents points that lie on the Seyfert/LINER division line (SLI$=0$). In agreement with the results of \cite{Ma21}, the nucleus and ionization bicone are dominated by Seyfert-like activity. Surrounding it, a thin LINER ``cocoon'' ($\sim 100$~pc) is characterized by intermediate SLI values, transitioning further into LINER-type emission with SLI$<-0.2$.

\subsection{Metallicity diagnostics}
To calculate $\log {\rm (O/H)}$ we used the metallicity diagnostics presented in Z24 for AGN-type emission with { MAPPINGS~V} \citep{Sutherland18}, which use incident AGN spectral energy distribution generated based on {\it OXAF} \citep{Thomas16}, a simplified version of {\it OPTXAGNF} \citep{Done12,Jin12} for thin-disk accretion.

We applied two metallicity diagnostics, both strongly dependent on { the gas-phase metallicity,} $\log {\rm (O/H)}$, but not on other gas parameters such as the ionization parameter, $\log U$, gas pressure, and the peak energy of the AGN SED, $E_p$.  They are given by the following line ratios:
\begin{eqnarray}\label{eq1}
   R1  &&= \log R(N2,S2,{\rm H}\alpha) =  \nonumber\\
    &&=\log\bigg(\frac{[N~II]\lambda6584}{[S~II]\lambda\lambda6717,31}\bigg) \nonumber \\ 
    &&+ 0.264\log\bigg(\frac{[N~II]\lambda6584}{{\rm H}\alpha}\bigg) 
\end{eqnarray}
and 
\begin{eqnarray}\label{eq2}
   R2  &&= \log R(N2,S2,O3) =  \nonumber\\
    &&=0.9738\log\bigg(\frac{[N~II]\lambda6584}{[S~II]\lambda\lambda6717,31}\bigg)  \nonumber\\
    &&+0.047\log\bigg(\frac{[N~II]\lambda6584}{{\rm H}\alpha}\bigg) \nonumber \\ 
    && - 0.183\log\bigg(\frac{[O~III]\lambda5007}{{\rm H}\beta}\bigg).
\end{eqnarray}
{ Both diagnostics are based on ratios of lines very close in the wavelength, ensuring limited impact of the dust extinction on the results.   }

We constructed metallicity maps for two N/O-{\rm O/H} scaling relations \citep{Groves04,Dors17} { corresponding to the  galaxies' different star formation history}: ``low'', $\log({\rm N/O}) = \log(10^{-1.732} + 10^{\log({\rm O/H})+2.19})$, and ``high'', $\log({\rm N/O}) = 1.29\times(12+\log({\rm O/H})-11.84)$. {  We derived the $\log({\rm O/H})$ uncertainties using the standard error propagation formula, combining the individual filter measurement errors, $\sigma$, with the diagnostics’ RMS (see Table~4 in Z24).}

Depending on the origin of the LINER-type emission, its metallicity can be approximated either using diagnostic methods similar to those for AGN-type emission, or using diagnostics derived for shock-excited regions, in the cases of obscured AGN radiation and shock-driven emission, respectively \citep{Halpern83,Dopita95}. We use both approaches. We approximate LINER-type regions average metallicity by applying  the AGN-type diagnostics (Eq.~\ref{eq1},\ref{eq2}) for the photoionization excitation and in the shock-driven emission scenario (Zhu et al., in preparation).

\subsection{Metallicity maps}
Based on the SLI map we separated AGN- and LINER-type-dominated regions. For Seyfert-dominated regions, we calculated R1 and R2 values on a pixel-by-pixel basis and then inserted them into the theoretically derived metallicity diagnostics to calculate $\log {\rm(O/H)}$ (Tab.~4 in Z24). We adopted $\log(E_p/\textrm{keV})=-1.5$ and the $\log U$ values by Zhu et al. (in preparation), derived based on {\it MUSE}-observed [O~III]/[O~I] emission line ratios and Z24 line diagnostics.  { We confirmed that derived metallicities are insensitive to $E_p$ and $\log U$ by testing a range of values. This is consistent with Z24.} { The mean value of $12+\log {\rm(O/H)}$ uncertainties derived using the standard error propagation formula for uncorrelated variables is $\sim0.035$~dex. }

\begin{figure*}
    \centering
    \includegraphics[width=1\textwidth]{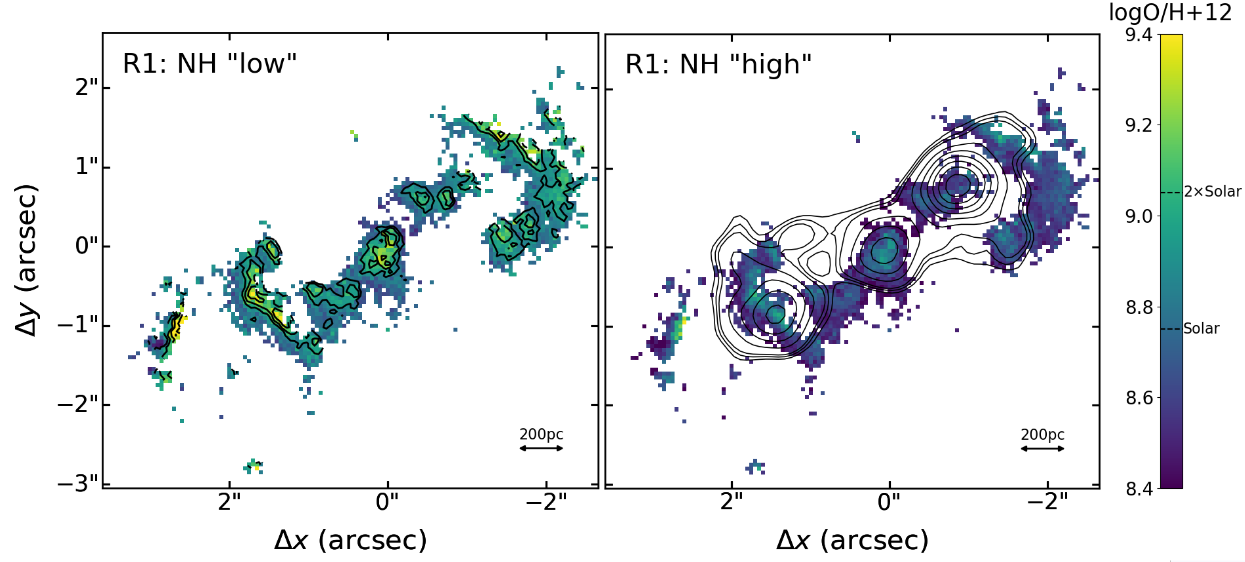}
    \includegraphics[width=1\textwidth]{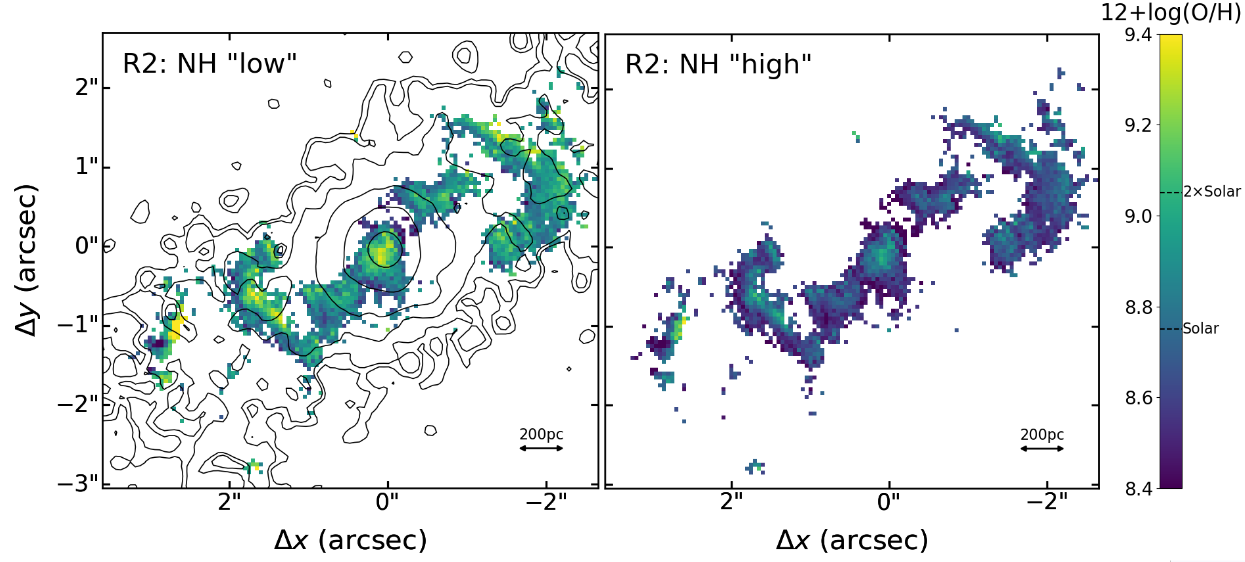}
    \caption{Metallicity ($\log {\rm (O/H)}+12$) maps calculated for AGN-type regions, based on R1 (upper row) and R2 (lower row) metallicity diagnostics and assuming low (left panels) and high (right panels) nitrogen scaling relations. Black contours trace SLI value on the top-left panel,  $6$~cm radio jet/radio-lobes emission and soft ($0.3$-$2.0$~keV) Chandra X-ray emission in bottom-left panel. Dashed lines on the color scale mark Solar, $2\times$Solar and $5\times$Solar abundances \citep[assuming $12 + \log {\rm( O/H)}= 8.75$ for the Solar value,][]{Bergemann21}. }
    \label{fig:metal}
\end{figure*}

\section{Results}\label{sec:results}

Fig.~\ref{fig:metal} shows the metallicity maps, represented as $12+\log({\rm O/H})$, for the $R1$ (upper panels of Fig.~\ref{fig:metal}) and $R2$ (lower panels of Fig.~\ref{fig:metal}) diagnostics. Both the ``low'' (left panels of Fig.~\ref{fig:metal}) and ``high'' (right panels of Fig.~\ref{fig:metal}) N/O–{\rm O/H} scalings consistently show metal enrichment in AGN-type regions. Surprisingly, we find enhanced, { super-solar } metallicity associated with the extended AGN excitation regions of Mrk~573. For both scaling relations some areas reach values up to $5\times$ Solar for the ``low'' N/O–{\rm O/H} scaling \citep[assuming $12+\log {\rm (O/H)} = 8.75$ for the Solar value,][]{Bergemann21}.  The $\log {\rm (O/H)}$ distribution within the bi-cone areas shows features on $\sim100$~pc scales. High metallicity appears to trace the { high SLI values}, radio emission at $6$~cm and the X-ray emission (shown as black contours in Fig.~\ref{fig:metal}).  {{ We note that other theoretical diagnostics are known to overestimate metallicities by $\sim 0.2$~dex compared to direct methods \citep{Dors20b}. The presence of regions with super-solar metallicity would not be affected by a similar offset in the Z24 diagnostic.}}

{ The metallicity of AGN-type regions is enhanced with respect to the LINER-type regions (blue in Fig.~\ref{fig:SLI}). The mean $\log{\rm(O/H)}$ of LINER regions is $\sim0.35$~dex lower for both the photoionized (R1/R2) and shock based diagnostics (Zhu et al. 2026, in preparation, Z26). The LINER-type points roughly follow the $12+\log{\rm(O/H)}$ and SLI correlation seen for the AGN-type points.  }

{ As reported by \cite{Riffel21}, emission within the inner kpc may be partially shock-induced, potentially biasing gas properties derived from photoionization diagnostics. However, our tests show that the results are robust to shock contributions in the bi-cone regions, with shock-based diagnostics (Z26) resulting in similar results.  }

\section{Discussion}\label{sec:discussion}

What is the origin of the enhanced metallicity associated with the AGN excited ISM (Fig.~\ref{fig:metal})? 

The metal enrichment can be either (1) created within the ionization bi-cones or (2) entrained by winds, outflows, or jets. Since there is no evidence of star formation within the ionization bi-cones \citep{Ma21}, the observed enrichment is unlikely to originate from in-situ stellar processes.

The central AGN regions, instead, are an appealing source of metals, as their broad-line regions (BLRs) are known to exhibit super-solar abundances \citep{Hamann99}. According to recent results, accretion disks are expected to be populated by stars, either formed in the accretion disk or captured \citep{Fabj25}. Metals created in the disk/torus would then have to be transported to scales of up to a few kiloparsecs. Below we briefly discuss two possible metal transportation mechanisms.

\subsection{Enrichment via winds and outflows}
The first possibility is that metals are carried outward by AGN-driven winds and outflows into the bi-cone regions. Recent {\it JWST} observations of massive outflows in radio-loud AGN \citep{Riffel26} provide evidence of the impact of AGN outflows on the bi-cone ISM. Other studies support AGN-driven outflows prevalence \citep[e.g.,][]{Deconto22,Revalski25}. 

{ In particular, outflows with velocities up to $\sim1000$~km~s$^{-1}$ and $\dot{M}_{obs}\sim 1~M_{\odot}$~yr$^{-1}$ mass outflow rates have been found up to $\sim1$~kpc from the Mrk~573 nucleus \citep{Revalski18}. Following the approach of \cite{Cano12}, we estimated the mass of the emitting gas in AGN-type regions as:}

\begin{equation}
M = 5.3\times10^7 \frac{L_{44}([O~III])}{n_{e3} Z} M_{\odot}=(3-5)\times10^5M_{\odot},
\end{equation}
with $Z$ denoting the oxygen abundance in Solar units, $n_{e3}$ the electron density of the gas in units of $10^3$~cm$^{-3}$, and $L_{44}([O~III])$ the [O~III] luminosity in units of $10^{44}$~erg~s$^{-1}$. { Based on  [S~II]$\lambda\lambda6716, 673$~\AA~fluxes measured by {\it MUSE}, we estimated $n_{e3}$ following \cite{Sanders16} method. } The mass of the gas is calculated on a pixel-by-pixel basis using the [O~III] emission, metallicity and $n_{e3}$ maps. Derived mass together with measured $\dot{M}_{obs}$ indicate timescales of $\sim3\times10^{5}$~yrs needed to displace metals from the AGN central regions to the NLR. These timescales match the time needed to reach $\sim1$~kpc radius at $1000$~km~s$^{-1}$ velocity. 

{ Can radiation driven winds move that much matter? The bolometric luminosity, $L_{bol}$, of Mrk~573 is $5.2\times10^{41}$~erg~s$^{-1}$ \citep{Gonzalez10}. For the outflow velocity, $v_w=1000$~km~s$^{-1}$  the mass rate of a radiation driven wind is: }
\begin{equation}
    \dot{M}_w = \frac{L_{bol}}{cv_w}\sim2.8M_{\odot}yr^{-1}.
\end{equation}
{ Assuming a covering factor of $\sim0.5$ the derived $\dot{M}_w$ is consistent with $\dot{M}_{obs}$. }

\subsection{Enrichment via Jets}

In the second scenario, jets play an important role in { detection of metal enrichment.}
This may be the case in Mrk~573, since there is radio emission in the metal-enriched areas (Fig.~\ref{fig:metal}, right panels).

Jet-associated metal displacement has been seen before, e.g., in the Hydra~A galaxy \citep{Kirkpatrick09}.  Theoretical and numerical considerations, although in the context of galaxy clusters, show that jets may drag up to $\sim10^{8}$--$10^{9}$~M$_{\odot}$ of gas \citep{DeYoung86,Simionescu09,Pope10}, well above our estimated mass of the metal-enriched gas. 

{ The jet kinetic power, $P_j$ can be estimated as $\sim4\times10^{41}$~erg~s$^{-1}$ based on its luminosity at $1.4$~GHz,$L_{1.4}$ \citep[see][]{Cavagnolo10}. We adopted $L_{1.4}=2.48\times10^{-2}$~Jy \citep[][]{Condon98}.  The kinetic luminosity of the $M\sim5\times10^5 M_{\odot}$ gas moving with $v_m=1000$~km~s$^{-1}$ velocity can be estimated as $\sim2\times10^{41}$~erg~s$^{-1}$, therefore based on the energetics alone, jet could move that amount of gas.}

{ More importantly, } jet-induced shocks in Mrk~573 may play an important role in the metal-enrichment if it originates from dust outflows, such as those found in high-redshift galaxies \citep[][]{Vayner25}. Metal-rich dust can be pushed from the AGN torus into the bi-cone areas, where shocks and photoionization may destroy it and release the metals in the ISM, see the ``engulfed cloud'' model, invoked as one of the scenarios explaining the absorption features in Compact Symmetric Objects \citep[CSOs, ][]{Begelman99}.  This scenario could explain the correlation between $\log {\rm (O/H)}$ and SLI, as high SLI values seem to correlate with the presence of fast shocks \citep{Krol25}.

\section{Summary and Conclusion}\label{sec:summary}
In this paper we report metal enrichment in the AGN-type bi-cone areas of Mrk~573. Enrichment is seen in $\log {\rm (O/H)}$ derived with both theoretical metallicity diagnostics and with {\rm N/O}–{\rm O/H} scaling relations, with $\log {\rm (O/H)}$ values reaching up to five times the Solar value for the ``low'' {\rm N/O}–{\rm O/H} scaling relation. { Considering both the AGN photo-ionized and shock origin for LINER-type regions,} the derived metallicity values are higher in AGN-type regions than in LINER-type regions.

{ The  $\log {\rm (O/H)}$ values trace the SLI values, soft X-ray emission and $6$~cm radio emission}, 
which indicates a strong link between the metal enrichment and the AGN-type emission mechanism, and we discussed potential avenues of metal-enriched gas transportation from the active nucleus to the NLR region: (1) metal-rich outflows, jet/radio-lobe-driven (2) metal displacement, and the possible role of associated shocks in releasing metals bound in dust outflows from the torus. 

The distinction between these mechanisms can be made through detailed studies of other objects. A similar metallicity enrichment is seen in other seven nearby Seyfert~2 galaxies. Future analysis will include application of theoretical diagnostics to {\it HST} and {\it MUSE } data, and comparison of a larger sample of Seyfert 2 galaxies with X-ray, radio and IR emission  (Król, et al. 2026 in preparation, Zhu et al. 2026 in preparation).

\begin{acknowledgments}

\end{acknowledgments}

\facilities{HST, CXO, VLA, MUSE}

\software{Astropy \citep{Astropy1,Astropy2,Astropy3}
          }

\bibliography{sample701}{}

@ARTICLE{Allen08,
       author = {{Allen}, Mark G. and {Groves}, Brent A. and {Dopita}, Michael A. and {Sutherland}, Ralph S. and {Kewley}, Lisa J.},
        title = "{The MAPPINGS III Library of Fast Radiative Shock Models}",
      journal = {\apjs},
         year = 2008,
        month = sep,
       volume = {178},
       number = {1},
        pages = {20-55},
          doi = {10.1086/589652},
archivePrefix = {arXiv},
       eprint = {0805.0204},
 primaryClass = {astro-ph},
       adsurl = {https://ui.adsabs.harvard.edu/abs/2008ApJS..178...20A}
}

@ARTICLE{Amiri24,
       author = {{Amiri}, A. and {Knapen}, J.~H. and {Comer{\'o}n}, S. and {Marconi}, A. and {Lehmer}, B.~D.},
        title = "{Gas-phase metallicity for the Seyfert galaxy NGC 7130}",
      journal = {\aap},
         year = 2024,
        month = sep,
       volume = {689},
          eid = {A193},
        pages = {A193},
          doi = {10.1051/0004-6361/202451168},
archivePrefix = {arXiv},
       eprint = {2407.12158},
 primaryClass = {astro-ph.GA},
       adsurl = {https://ui.adsabs.harvard.edu/abs/2024A&A...689A.193A}
}

@ARTICLE{Amiri25,
       author = {{Amiri}, Amirnezam and {Knapen}, Johan H. and {Lehmer}, Bret D. and {Khoram}, Amirhossein},
        title = "{Negative gas-phase metallicity gradients in the narrow line region and galactic disc of local AGN-host galaxies}",
      journal = {arXiv e-prints},
         year = 2025,
        month = sep,
          eid = {arXiv:2509.12156},
        pages = {arXiv:2509.12156},
          doi = {10.48550/arXiv.2509.12156},
archivePrefix = {arXiv},
       eprint = {2509.12156},
 primaryClass = {astro-ph.GA},
       adsurl = {https://ui.adsabs.harvard.edu/abs/2025arXiv250912156A}
}

@ARTICLE{Armah24,
       author = {{Armah}, Mark and {Riffel}, Rog{\'e}rio and {Dahmer-Hahn}, L.~G. and {Davies}, R.~I. and {Dors}, O.~L. and {Kakkad}, Darshan and {Riffel}, Rogemar A. and {Rodr{\'\i}guez-Ardila}, A. and {Ruschel-Dutra}, D. and {Storchi-Bergmann}, T.},
        title = "{Spatially resolved gas-phase metallicity in Seyfert galaxies}",
      journal = {\mnras},
         year = 2024,
        month = nov,
       volume = {534},
       number = {3},
        pages = {2723-2757},
          doi = {10.1093/mnras/stae2263},
archivePrefix = {arXiv},
       eprint = {2409.20465},
 primaryClass = {astro-ph.GA},
       adsurl = {https://ui.adsabs.harvard.edu/abs/2024MNRAS.534.2723A}
}

@ARTICLE{Astropy1,
       author = {{Astropy Collaboration} and {Price-Whelan}, Adrian M. and {Lim}, Pey Lian and {Earl}, Nicholas and {Starkman}, Nathaniel and {Bradley}, Larry and {Shupe}, David L. and {Patil}, Aarya A. and {Corrales}, Lia and {Brasseur}, C.~E. and {N{\"o}the}, Maximilian and {Donath}, Axel and {Tollerud}, Erik and {Morris}, Brett M. and {Ginsburg}, Adam and {Vaher}, Eero and {Weaver}, Benjamin A. and {Tocknell}, James and {Jamieson}, William and {van Kerkwijk}, Marten H. and {Robitaille}, Thomas P. and {Merry}, Bruce and {Bachetti}, Matteo and {G{\"u}nther}, H. Moritz and {Aldcroft}, Thomas L. and {Alvarado-Montes}, Jaime A. and {Archibald}, Anne M. and {B{\'o}di}, Attila and {Bapat}, Shreyas and {Barentsen}, Geert and {Baz{\'a}n}, Juanjo and {Biswas}, Manish and {Boquien}, M{\'e}d{\'e}ric and {Burke}, D.~J. and {Cara}, Daria and {Cara}, Mihai and {Conroy}, Kyle E. and {Conseil}, Simon and {Craig}, Matthew W. and {Cross}, Robert M. and {Cruz}, Kelle L. and {D'Eugenio}, Francesco and {Dencheva}, Nadia and {Devillepoix}, Hadrien A.~R. and {Dietrich}, J{\"o}rg P. and {Eigenbrot}, Arthur Davis and {Erben}, Thomas and {Ferreira}, Leonardo and {Foreman-Mackey}, Daniel and {Fox}, Ryan and {Freij}, Nabil and {Garg}, Suyog and {Geda}, Robel and {Glattly}, Lauren and {Gondhalekar}, Yash and {Gordon}, Karl D. and {Grant}, David and {Greenfield}, Perry and {Groener}, Austen M. and {Guest}, Steve and {Gurovich}, Sebastian and {Handberg}, Rasmus and {Hart}, Akeem and {Hatfield-Dodds}, Zac and {Homeier}, Derek and {Hosseinzadeh}, Griffin and {Jenness}, Tim and {Jones}, Craig K. and {Joseph}, Prajwel and {Kalmbach}, J. Bryce and {Karamehmetoglu}, Emir and {Ka{\l}uszy{\'n}ski}, Miko{\l}aj and {Kelley}, Michael S.~P. and {Kern}, Nicholas and {Kerzendorf}, Wolfgang E. and {Koch}, Eric W. and {Kulumani}, Shankar and {Lee}, Antony and {Ly}, Chun and {Ma}, Zhiyuan and {MacBride}, Conor and {Maljaars}, Jakob M. and {Muna}, Demitri and {Murphy}, N.~A. and {Norman}, Henrik and {O'Steen}, Richard and {Oman}, Kyle A. and {Pacifici}, Camilla and {Pascual}, Sergio and {Pascual-Granado}, J. and {Patil}, Rohit R. and {Perren}, Gabriel I. and {Pickering}, Timothy E. and {Rastogi}, Tanuj and {Roulston}, Benjamin R. and {Ryan}, Daniel F. and {Rykoff}, Eli S. and {Sabater}, Jose and {Sakurikar}, Parikshit and {Salgado}, Jes{\'u}s and {Sanghi}, Aniket and {Saunders}, Nicholas and {Savchenko}, Volodymyr and {Schwardt}, Ludwig and {Seifert-Eckert}, Michael and {Shih}, Albert Y. and {Jain}, Anany Shrey and {Shukla}, Gyanendra and {Sick}, Jonathan and {Simpson}, Chris and {Singanamalla}, Sudheesh and {Singer}, Leo P. and {Singhal}, Jaladh and {Sinha}, Manodeep and {Sip{\H{o}}cz}, Brigitta M. and {Spitler}, Lee R. and {Stansby}, David and {Streicher}, Ole and {{\v{S}}umak}, Jani and {Swinbank}, John D. and {Taranu}, Dan S. and {Tewary}, Nikita and {Tremblay}, Grant R. and {de Val-Borro}, Miguel and {Van Kooten}, Samuel J. and {Vasovi{\'c}}, Zlatan and {Verma}, Shresth and {de Miranda Cardoso}, Jos{\'e} Vin{\'\i}cius and {Williams}, Peter K.~G. and {Wilson}, Tom J. and {Winkel}, Benjamin and {Wood-Vasey}, W.~M. and {Xue}, Rui and {Yoachim}, Peter and {Zhang}, Chen and {Zonca}, Andrea and {Astropy Project Contributors}},
        title = "{The Astropy Project: Sustaining and Growing a Community-oriented Open-source Project and the Latest Major Release (v5.0) of the Core Package}",
      journal = {\apj},
         year = 2022,
        month = aug,
       volume = {935},
       number = {2},
          eid = {167},
        pages = {167},
          doi = {10.3847/1538-4357/ac7c74},
archivePrefix = {arXiv},
       eprint = {2206.14220},
 primaryClass = {astro-ph.IM},
       adsurl = {https://ui.adsabs.harvard.edu/abs/2022ApJ...935..167A}
}

@ARTICLE{Astropy2,
       author = {{Astropy Collaboration} and {Price-Whelan}, A.~M. and {Sip{\H{o}}cz}, B.~M. and {G{\"u}nther}, H.~M. and {Lim}, P.~L. and {Crawford}, S.~M. and {Conseil}, S. and {Shupe}, D.~L. and {Craig}, M.~W. and {Dencheva}, N. and {Ginsburg}, A. and {VanderPlas}, J.~T. and {Bradley}, L.~D. and {P{\'e}rez-Su{\'a}rez}, D. and {de Val-Borro}, M. and {Aldcroft}, T.~L. and {Cruz}, K.~L. and {Robitaille}, T.~P. and {Tollerud}, E.~J. and {Ardelean}, C. and {Babej}, T. and {Bach}, Y.~P. and {Bachetti}, M. and {Bakanov}, A.~V. and {Bamford}, S.~P. and {Barentsen}, G. and {Barmby}, P. and {Baumbach}, A. and {Berry}, K.~L. and {Biscani}, F. and {Boquien}, M. and {Bostroem}, K.~A. and {Bouma}, L.~G. and {Brammer}, G.~B. and {Bray}, E.~M. and {Breytenbach}, H. and {Buddelmeijer}, H. and {Burke}, D.~J. and {Calderone}, G. and {Cano Rodr{\'\i}guez}, J.~L. and {Cara}, M. and {Cardoso}, J.~V.~M. and {Cheedella}, S. and {Copin}, Y. and {Corrales}, L. and {Crichton}, D. and {D'Avella}, D. and {Deil}, C. and {Depagne}, {\'E}. and {Dietrich}, J.~P. and {Donath}, A. and {Droettboom}, M. and {Earl}, N. and {Erben}, T. and {Fabbro}, S. and {Ferreira}, L.~A. and {Finethy}, T. and {Fox}, R.~T. and {Garrison}, L.~H. and {Gibbons}, S.~L.~J. and {Goldstein}, D.~A. and {Gommers}, R. and {Greco}, J.~P. and {Greenfield}, P. and {Groener}, A.~M. and {Grollier}, F. and {Hagen}, A. and {Hirst}, P. and {Homeier}, D. and {Horton}, A.~J. and {Hosseinzadeh}, G. and {Hu}, L. and {Hunkeler}, J.~S. and {Ivezi{\'c}}, {\v{Z}}. and {Jain}, A. and {Jenness}, T. and {Kanarek}, G. and {Kendrew}, S. and {Kern}, N.~S. and {Kerzendorf}, W.~E. and {Khvalko}, A. and {King}, J. and {Kirkby}, D. and {Kulkarni}, A.~M. and {Kumar}, A. and {Lee}, A. and {Lenz}, D. and {Littlefair}, S.~P. and {Ma}, Z. and {Macleod}, D.~M. and {Mastropietro}, M. and {McCully}, C. and {Montagnac}, S. and {Morris}, B.~M. and {Mueller}, M. and {Mumford}, S.~J. and {Muna}, D. and {Murphy}, N.~A. and {Nelson}, S. and {Nguyen}, G.~H. and {Ninan}, J.~P. and {N{\"o}the}, M. and {Ogaz}, S. and {Oh}, S. and {Parejko}, J.~K. and {Parley}, N. and {Pascual}, S. and {Patil}, R. and {Patil}, A.~A. and {Plunkett}, A.~L. and {Prochaska}, J.~X. and {Rastogi}, T. and {Reddy Janga}, V. and {Sabater}, J. and {Sakurikar}, P. and {Seifert}, M. and {Sherbert}, L.~E. and {Sherwood-Taylor}, H. and {Shih}, A.~Y. and {Sick}, J. and {Silbiger}, M.~T. and {Singanamalla}, S. and {Singer}, L.~P. and {Sladen}, P.~H. and {Sooley}, K.~A. and {Sornarajah}, S. and {Streicher}, O. and {Teuben}, P. and {Thomas}, S.~W. and {Tremblay}, G.~R. and {Turner}, J.~E.~H. and {Terr{\'o}n}, V. and {van Kerkwijk}, M.~H. and {de la Vega}, A. and {Watkins}, L.~L. and {Weaver}, B.~A. and {Whitmore}, J.~B. and {Woillez}, J. and {Zabalza}, V. and {Astropy Contributors}},
        title = "{The Astropy Project: Building an Open-science Project and Status of the v2.0 Core Package}",
      journal = {\aj},
         year = 2018,
        month = sep,
       volume = {156},
       number = {3},
          eid = {123},
        pages = {123},
          doi = {10.3847/1538-3881/aabc4f},
archivePrefix = {arXiv},
       eprint = {1801.02634},
 primaryClass = {astro-ph.IM},
       adsurl = {https://ui.adsabs.harvard.edu/abs/2018AJ....156..123A}
}

@ARTICLE{Astropy3,
       author = {{Astropy Collaboration} and {Robitaille}, Thomas P. and {Tollerud}, Erik J. and {Greenfield}, Perry and {Droettboom}, Michael and {Bray}, Erik and {Aldcroft}, Tom and {Davis}, Matt and {Ginsburg}, Adam and {Price-Whelan}, Adrian M. and {Kerzendorf}, Wolfgang E. and {Conley}, Alexander and {Crighton}, Neil and {Barbary}, Kyle and {Muna}, Demitri and {Ferguson}, Henry and {Grollier}, Fr{\'e}d{\'e}ric and {Parikh}, Madhura M. and {Nair}, Prasanth H. and {Unther}, Hans M. and {Deil}, Christoph and {Woillez}, Julien and {Conseil}, Simon and {Kramer}, Roban and {Turner}, James E.~H. and {Singer}, Leo and {Fox}, Ryan and {Weaver}, Benjamin A. and {Zabalza}, Victor and {Edwards}, Zachary I. and {Azalee Bostroem}, K. and {Burke}, D.~J. and {Casey}, Andrew R. and {Crawford}, Steven M. and {Dencheva}, Nadia and {Ely}, Justin and {Jenness}, Tim and {Labrie}, Kathleen and {Lim}, Pey Lian and {Pierfederici}, Francesco and {Pontzen}, Andrew and {Ptak}, Andy and {Refsdal}, Brian and {Servillat}, Mathieu and {Streicher}, Ole},
        title = "{Astropy: A community Python package for astronomy}",
      journal = {\aap},
         year = 2013,
        month = oct,
       volume = {558},
          eid = {A33},
        pages = {A33},
          doi = {10.1051/0004-6361/201322068},
archivePrefix = {arXiv},
       eprint = {1307.6212},
 primaryClass = {astro-ph.IM},
       adsurl = {https://ui.adsabs.harvard.edu/abs/2013A&A...558A..33A}
}

@INPROCEEDINGS{Bacon10,
       author = {{Bacon}, R. and {Accardo}, M. and {Adjali}, L. and {Anwand}, H. and {Bauer}, S. and {Biswas}, I. and {Blaizot}, J. and {Boudon}, D. and {Brau-Nogue}, S. and {Brinchmann}, J. and {Caillier}, P. and {Capoani}, L. and {Carollo}, C.~M. and {Contini}, T. and {Couderc}, P. and {Daguis{\'e}}, E. and {Deiries}, S. and {Delabre}, B. and {Dreizler}, S. and {Dubois}, J. and {Dupieux}, M. and {Dupuy}, C. and {Emsellem}, E. and {Fechner}, T. and {Fleischmann}, A. and {Fran{\c{c}}ois}, M. and {Gallou}, G. and {Gharsa}, T. and {Glindemann}, A. and {Gojak}, D. and {Guiderdoni}, B. and {Hansali}, G. and {Hahn}, T. and {Jarno}, A. and {Kelz}, A. and {Koehler}, C. and {Kosmalski}, J. and {Laurent}, F. and {Le Floch}, M. and {Lilly}, S.~J. and {Lizon}, J.-L. and {Loupias}, M. and {Manescau}, A. and {Monstein}, C. and {Nicklas}, H. and {Olaya}, J.-C. and {Pares}, L. and {Pasquini}, L. and {P{\'e}contal-Rousset}, A. and {Pell{\'o}}, R. and {Petit}, C. and {Popow}, E. and {Reiss}, R. and {Remillieux}, A. and {Renault}, E. and {Roth}, M. and {Rupprecht}, G. and {Serre}, D. and {Schaye}, J. and {Soucail}, G. and {Steinmetz}, M. and {Streicher}, O. and {Stuik}, R. and {Valentin}, H. and {Vernet}, J. and {Weilbacher}, P. and {Wisotzki}, L. and {Yerle}, N.},
        title = "{The MUSE second-generation VLT instrument}",
    booktitle = {Ground-based and Airborne Instrumentation for Astronomy III},
         year = 2010,
       editor = {{McLean}, Ian S. and {Ramsay}, Suzanne K. and {Takami}, Hideki},
       series = {Society of Photo-Optical Instrumentation Engineers (SPIE) Conference Series},
       volume = {7735},
        month = jul,
          eid = {773508},
        pages = {773508},
          doi = {10.1117/12.856027},
archivePrefix = {arXiv},
       eprint = {2211.16795},
 primaryClass = {astro-ph.IM},
       adsurl = {https://ui.adsabs.harvard.edu/abs/2010SPIE.7735E..08B}
}

@ARTICLE{Baldwin81,
       author = {{Baldwin}, J.~A. and {Phillips}, M.~M. and {Terlevich}, R.},
        title = "{Classification parameters for the emission-line spectra of extragalactic objects.}",
      journal = {\pasp},
         year = 1981,
        month = feb,
       volume = {93},
        pages = {5-19},
          doi = {10.1086/130766},
       adsurl = {https://ui.adsabs.harvard.edu/abs/1981PASP...93....5B}
}

@INPROCEEDINGS{Begelman99,
       author = {{Begelman}, Mitchell C.},
        title = "{Young radio galaxies and their environments}",
    booktitle = {The Most Distant Radio Galaxies},
         year = 1999,
       editor = {{R{\"o}ttgering}, H.~J.~A. and {Best}, P.~N. and {Lehnert}, M.~D.},
        month = jan,
        pages = {173},
       adsurl = {https://ui.adsabs.harvard.edu/abs/1999mdrg.conf..173B}
}

@ARTICLE{Bergemann21,
       author = {{Bergemann}, Maria and {Hoppe}, Richard and {Semenova}, Ekaterina and {Carlsson}, Mats and {Yakovleva}, Svetlana A. and {Voronov}, Yaroslav V. and {Bautista}, Manuel and {Nemer}, Ahmad and {Belyaev}, Andrey K. and {Leenaarts}, Jorrit and {Mashonkina}, Lyudmila and {Reiners}, Ansgar and {Ellwarth}, Monika},
        title = "{Solar oxygen abundance}",
      journal = {\mnras},
         year = 2021,
        month = dec,
       volume = {508},
       number = {2},
        pages = {2236-2253},
          doi = {10.1093/mnras/stab2160},
archivePrefix = {arXiv},
       eprint = {2109.01143},
 primaryClass = {astro-ph.SR},
       adsurl = {https://ui.adsabs.harvard.edu/abs/2021MNRAS.508.2236B}
}

@ARTICLE{Bianchi10,
       author = {{Bianchi}, Stefano and {Chiaberge}, Marco and {Evans}, Daniel A. and {Guainazzi}, Matteo and {Baldi}, Ranieri D. and {Matt}, Giorgio and {Piconcelli}, Enrico},
        title = "{High-resolution X-ray spectroscopy and imaging of Mrk 573}",
      journal = {\mnras},
         year = 2010,
        month = jun,
       volume = {405},
       number = {1},
        pages = {553-563},
          doi = {10.1111/j.1365-2966.2010.16475.x},
archivePrefix = {arXiv},
       eprint = {1002.0800},
 primaryClass = {astro-ph.GA},
       adsurl = {https://ui.adsabs.harvard.edu/abs/2010MNRAS.405..553B}
}

@ARTICLE{Binette94,
       author = {{Binette}, L. and {Magris}, C.~G. and {Stasi{\'n}ska}, G. and {Bruzual}, A.~G.},
        title = "{Photoionization in elliptical galaxies by old stars.}",
      journal = {\aap},
         year = 1994,
        month = dec,
       volume = {292},
        pages = {13-19},
       adsurl = {https://ui.adsabs.harvard.edu/abs/1994A&A...292...13B}
}

@ARTICLE{Bittner19,
       author = {{Bittner}, A. and {Falc{\'o}n-Barroso}, J. and {Nedelchev}, B. and {Dorta}, A. and {Gadotti}, D.~A. and {Sarzi}, M. and {Molaeinezhad}, A. and {Iodice}, E. and {Rosado-Belza}, D. and {de Lorenzo-C{\'a}ceres}, A. and {Fragkoudi}, F. and {Gal{\'a}n-de Anta}, P.~M. and {Husemann}, B. and {M{\'e}ndez-Abreu}, J. and {Neumann}, J. and {Pinna}, F. and {Querejeta}, M. and {S{\'a}nchez-Bl{\'a}zquez}, P. and {Seidel}, M.~K.},
        title = "{The GIST pipeline: A multi-purpose tool for the analysis and visualisation of (integral-field) spectroscopic data}",
      journal = {\aap},
         year = 2019,
        month = aug,
       volume = {628},
          eid = {A117},
        pages = {A117},
          doi = {10.1051/0004-6361/201935829},
archivePrefix = {arXiv},
       eprint = {1906.04746},
 primaryClass = {astro-ph.GA},
       adsurl = {https://ui.adsabs.harvard.edu/abs/2019A&A...628A.117B}
}

@ARTICLE{Cano12,
       author = {{Cano-D{\'\i}az}, M. and {Maiolino}, R. and {Marconi}, A. and {Netzer}, H. and {Shemmer}, O. and {Cresci}, G.},
        title = "{Observational evidence of quasar feedback quenching star formation at high redshift}",
      journal = {\aap},
         year = 2012,
        month = jan,
       volume = {537},
          eid = {L8},
        pages = {L8},
          doi = {10.1051/0004-6361/201118358},
archivePrefix = {arXiv},
       eprint = {1112.3071},
 primaryClass = {astro-ph.CO},
       adsurl = {https://ui.adsabs.harvard.edu/abs/2012A&A...537L...8C}
}

@ARTICLE{Cappellari04,
       author = {{Cappellari}, Michele and {Emsellem}, Eric},
        title = "{Parametric Recovery of Line-of-Sight Velocity Distributions from Absorption-Line Spectra of Galaxies via Penalized Likelihood}",
      journal = {\pasp},
         year = 2004,
        month = feb,
       volume = {116},
       number = {816},
        pages = {138-147},
          doi = {10.1086/381875},
archivePrefix = {arXiv},
       eprint = {astro-ph/0312201},
 primaryClass = {astro-ph},
       adsurl = {https://ui.adsabs.harvard.edu/abs/2004PASP..116..138C}
}

@ARTICLE{Cappellari17,
       author = {{Cappellari}, Michele},
        title = "{Improving the full spectrum fitting method: accurate convolution with Gauss-Hermite functions}",
      journal = {\mnras},
         year = 2017,
        month = apr,
       volume = {466},
       number = {1},
        pages = {798-811},
          doi = {10.1093/mnras/stw3020},
archivePrefix = {arXiv},
       eprint = {1607.08538},
 primaryClass = {astro-ph.GA},
       adsurl = {https://ui.adsabs.harvard.edu/abs/2017MNRAS.466..798C}
}

@ARTICLE{Castro17,
       author = {{Castro}, C.~S. and {Dors}, O.~L. and {Cardaci}, M.~V. and {H{\"a}gele}, G.~F.},
        title = "{New metallicity calibration for Seyfert 2 galaxies based on the N2O2 index}",
      journal = {\mnras},
         year = 2017,
        month = may,
       volume = {467},
       number = {2},
        pages = {1507-1514},
          doi = {10.1093/mnras/stx150},
archivePrefix = {arXiv},
       eprint = {1701.04997},
 primaryClass = {astro-ph.GA},
       adsurl = {https://ui.adsabs.harvard.edu/abs/2017MNRAS.467.1507C}
}

@ARTICLE{Cavagnolo10,
       author = {{Cavagnolo}, K.~W. and {McNamara}, B.~R. and {Nulsen}, P.~E.~J. and {Carilli}, C.~L. and {Jones}, C. and {B{\^\i}rzan}, L.},
        title = "{A Relationship Between AGN Jet Power and Radio Power}",
      journal = {\apj},
         year = 2010,
        month = sep,
       volume = {720},
       number = {2},
        pages = {1066-1072},
          doi = {10.1088/0004-637X/720/2/1066},
archivePrefix = {arXiv},
       eprint = {1006.5699},
 primaryClass = {astro-ph.CO},
       adsurl = {https://ui.adsabs.harvard.edu/abs/2010ApJ...720.1066C}
}

@ARTICLE{Carvalho20,
       author = {{Carvalho}, S.~P. and {Dors}, O.~L. and {Cardaci}, M.~V. and {H{\"a}gele}, G.~F. and {Krabbe}, A.~C. and {P{\'e}rez-Montero}, E. and {Monteiro}, A.~F. and {Armah}, M. and {Freitas-Lemes}, P.},
        title = "{Chemical abundances of Seyfert 2 AGNs - II. N2 metallicity calibration based on SDSS}",
      journal = {\mnras},
         year = 2020,
        month = mar,
       volume = {492},
       number = {4},
        pages = {5675-5683},
          doi = {10.1093/mnras/staa193},
archivePrefix = {arXiv},
       eprint = {2001.07126},
 primaryClass = {astro-ph.GA},
       adsurl = {https://ui.adsabs.harvard.edu/abs/2020MNRAS.492.5675C}
}

@ARTICLE{Condon98,
       author = {{Condon}, J.~J. and {Cotton}, W.~D. and {Greisen}, E.~W. and {Yin}, Q.~F. and {Perley}, R.~A. and {Taylor}, G.~B. and {Broderick}, J.~J.},
        title = "{The NRAO VLA Sky Survey}",
      journal = {\aj},
         year = 1998,
        month = may,
       volume = {115},
       number = {5},
        pages = {1693-1716},
          doi = {10.1086/300337},
       adsurl = {https://ui.adsabs.harvard.edu/abs/1998AJ....115.1693C}
}

@ARTICLE{Deconto22,
       author = {{Deconto-Machado}, A. and {Riffel}, R.~A. and {Ilha}, G.~S. and {Rembold}, S.~B. and {Storchi-Bergmann}, T. and {Riffel}, R. and {Schimoia}, J.~S. and {Schneider}, D.~P. and {Bizyaev}, D. and {Feng}, S. and {Wylezalek}, D. and {da Costa}, L.~N. and {do Nascimento}, J.~C. and {Maia}, M.~A.~G.},
        title = "{Ionised gas kinematics in MaNGA AGN. Extents of the narrow-line and kinematically disturbed regions}",
      journal = {\aap},
         year = 2022,
        month = mar,
       volume = {659},
          eid = {A131},
        pages = {A131},
          doi = {10.1051/0004-6361/202140613},
archivePrefix = {arXiv},
       eprint = {2201.01690},
 primaryClass = {astro-ph.GA},
       adsurl = {https://ui.adsabs.harvard.edu/abs/2022A&A...659A.131D}
}

@ARTICLE{DeYoung86,
       author = {{De Young}, D.~S.},
        title = "{Mass Entrainment in Astrophysical Jets}",
      journal = {\apj},
         year = 1986,
        month = aug,
       volume = {307},
        pages = {62},
          doi = {10.1086/164393},
       adsurl = {https://ui.adsabs.harvard.edu/abs/1986ApJ...307...62D}
}

@ARTICLE{Done12,
       author = {{Done}, Chris and {Davis}, S.~W. and {Jin}, C. and {Blaes}, O. and {Ward}, M.},
        title = "{Intrinsic disc emission and the soft X-ray excess in active galactic nuclei}",
      journal = {\mnras},
         year = 2012,
        month = mar,
       volume = {420},
       number = {3},
        pages = {1848-1860},
          doi = {10.1111/j.1365-2966.2011.19779.x},
archivePrefix = {arXiv},
       eprint = {1107.5429},
 primaryClass = {astro-ph.HE},
       adsurl = {https://ui.adsabs.harvard.edu/abs/2012MNRAS.420.1848D}
}

@ARTICLE{Dopita95,
       author = {{Dopita}, Michael A. and {Sutherland}, Ralph S.},
        title = "{Spectral Signatures of Fast Shocks. II. Optical Diagnostic Diagrams}",
      journal = {\apj},
         year = 1995,
        month = dec,
       volume = {455},
        pages = {468},
          doi = {10.1086/176596},
       adsurl = {https://ui.adsabs.harvard.edu/abs/1995ApJ...455..468D}
}

@ARTICLE{Dors17,
       author = {{Dors}, Jr., O.~L. and {Arellano-C{\'o}rdova}, K.~Z. and {Cardaci}, M.~V. and {H{\"a}gele}, G.~F.},
        title = "{New quantitative nitrogen abundance estimations in a sample of Seyfert 2 active galactic nuclei}",
      journal = {\mnras},
         year = 2017,
        month = jun,
       volume = {468},
       number = {1},
        pages = {L113-L117},
          doi = {10.1093/mnrasl/slx036},
archivePrefix = {arXiv},
       eprint = {1703.03250},
 primaryClass = {astro-ph.GA},
       adsurl = {https://ui.adsabs.harvard.edu/abs/2017MNRAS.468L.113D}
}

@ARTICLE{Dors20b,
       author = {{Dors}, O.~L. and {Maiolino}, R. and {Cardaci}, M.~V. and {H{\"a}gele}, G.~F. and {Krabbe}, A.~C. and {P{\'e}rez-Montero}, E. and {Armah}, M.},
        title = "{Chemical abundances of Seyfert 2 AGNs - III. Reducing the oxygen abundance discrepancy}",
      journal = {\mnras},
         year = 2020,
        month = aug,
       volume = {496},
       number = {3},
        pages = {3209-3221},
          doi = {10.1093/mnras/staa1781},
archivePrefix = {arXiv},
       eprint = {2006.09152},
 primaryClass = {astro-ph.GA},
       adsurl = {https://ui.adsabs.harvard.edu/abs/2020MNRAS.496.3209D}
}

@ARTICLE{Du14,
       author = {{Du}, Pu and {Wang}, Jian-Min and {Hu}, Chen and {Valls-Gabaud}, David and {Baldwin}, Jack A. and {Ge}, Jun-Qiang and {Xue}, Sui-Jian},
        title = "{Outflows from active galactic nuclei: the BLR-NLR metallicity correlation}",
      journal = {\mnras},
         year = 2014,
        month = mar,
       volume = {438},
       number = {4},
        pages = {2828-2838},
          doi = {10.1093/mnras/stt2386},
archivePrefix = {arXiv},
       eprint = {1312.3212},
 primaryClass = {astro-ph.GA},
       adsurl = {https://ui.adsabs.harvard.edu/abs/2014MNRAS.438.2828D}
}

@ARTICLE{Fabj25,
       author = {{Fabj}, Gaia and {Dittmann}, Alexander J. and {Cantiello}, Matteo and {Perna}, Rosalba and {Samsing}, Johan},
        title = "{Mapping the Outcomes of Stellar Evolution in the Disks of Active Galactic Nuclei}",
      journal = {\apj},
         year = 2025,
        month = mar,
       volume = {981},
       number = {1},
          eid = {16},
        pages = {16},
          doi = {10.3847/1538-4357/ada896},
archivePrefix = {arXiv},
       eprint = {2408.16050},
 primaryClass = {astro-ph.GA},
       adsurl = {https://ui.adsabs.harvard.edu/abs/2025ApJ...981...16F}
}
\bibliographystyle{aasjournalv7}

\end{document}